\documentclass[aps, prd, amsmath, amssymb, reprint, groupedaddress]{revtex4-1}

\usepackage{graphicx}% Include figure files
\usepackage{dcolumn}% Align table columns on decimal point
\usepackage{bm}% bold math
\usepackage{xcolor}
\usepackage{physics}

\begin{document}

\preprint{AIP/123-QED}

%\title[Sample Title]{Sample Title:\\with Forced Linebreak\footnote{Error!}}% Force line breaks with \\
%\thanks{Footnote to title of article.}

\title[]{Multiparametric Amplification and Qubit Measurement with a Kerr-free Josephson Ring Modulator}
%\thanks{Footnote to title of article.}

\author{T.-C. Chien, O. Lanes, C. Liu, X. Cao, P. Lu, S. Motz, G. Liu, D. Pekker, and M. Hatridge}
\email{Correspondence should be addressed to: hatridge@pitt.edu.}
\affiliation{Department of Physics and Astronomy, University of Pittsburgh, Pittsburgh, Pennsylvania 15260, USA}

\date{\today}% It is always \today, today,

\begin{abstract}

Josephson-junction based parametric amplifiers have become a ubiquitous component in superconducting quantum machines. Although parametric amplifiers regularly achieve near-quantum limited performance, they have many limitations, including low saturation powers, lack of directionality, and narrow bandwidth.  The first is believed to stem from the higher order Hamiltonian terms endemic to Josephson junction circuits, and the latter two are direct consequences of the nature of the parametric interactions which power them.  In this work, we attack both of these issues. First, we have designed a new, linearly shunted Josephson Ring Modulator (JRM) which nearly nulls  all 4th-order terms at a single flux bias point. Next, we achieve gain through a pair of balanced parametric drives. When applied separately, these drives produce phase-preserving gain~(G) and gainless photon conversion~(C), when applied together, the resultant amplifier (which we term GC) is a bi-directional, phase-sensitive transmission-only amplifier with a large, gain-independent bandwidth. Finally, we have also demonstrated the practical utility of the GC amplifier, as well as its' quantum efficiency, by using it to read out a superconducting transmon qubit.

%Valid PACS numbers may be entered using the \verb+\pacs{#1}+ command.
\end{abstract}

%\pacs{TBD}% PACS, the Physics and Astronomy
%\keywords{TBD}%Use showkeys class option if keyword

\maketitle

\section {Introduction}

In recent years, the potential of superconducting qubits as a quantum computing architecture has grown tremendously~\cite{Devoret2013, Barends2014}. This is in part due to quantum limited, Josephson-junction based parametric amplifiers (JPAs), which are able to enhance qubit signals while adding the minimum amount of noise and back-action, allowed by quantum mechanics~\cite{Caves1982}. In these devices, gain is achieved by intensely driving the non-linear Hamiltonian of a circuit containing one or more Josephson junctions to parametrically couple together one or more microwave modes~\cite{Roy2016}.  Paramps regularly enable experiments with single-shot, high-fidelity quantum non-demolition measurements of superconducting quantum bits~\cite{Vijay2011, Hatridge2013, Walter2017}. 

\begin{figure*}[htbp]
	\includegraphics[scale = 1]{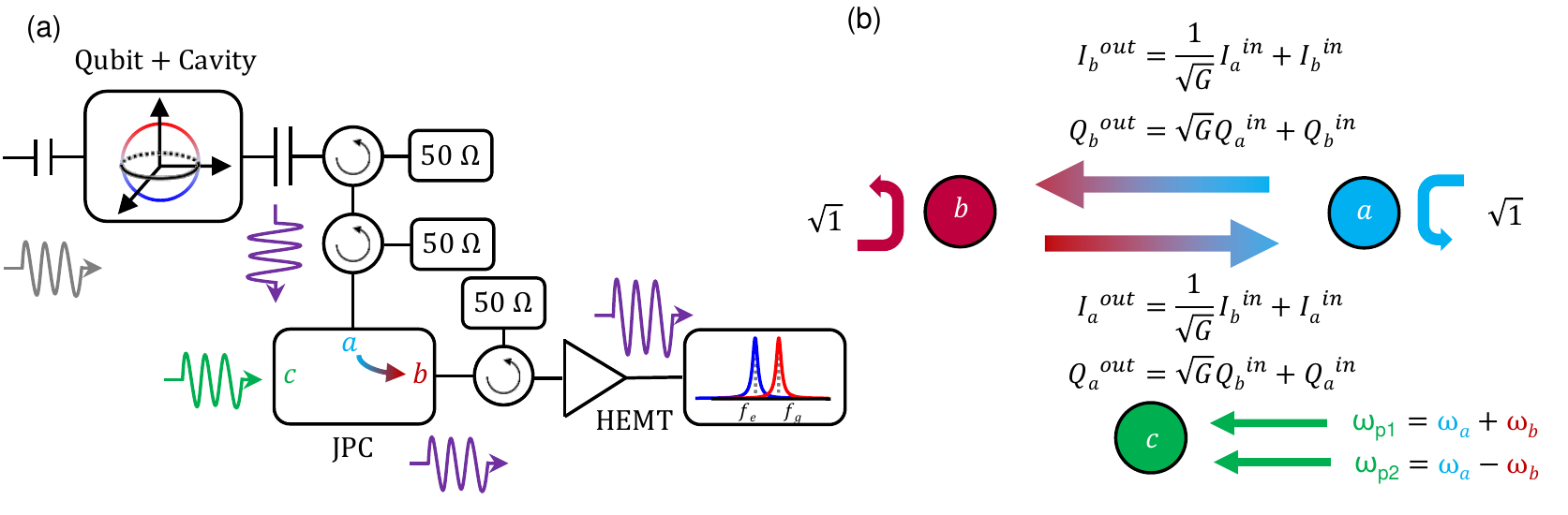}
		\caption{ \textbf{(a) Schematic of quantum measurement experiment.} A quantum limited amplifier, the Josephson parametric converter (JPC), is used in transmission to measure a transmon qubit mounted inside a 3D cavity. \textbf{ (b) Diagram of the `GC' scattering matrix. } The JPC has three microwave modes (modes $a$, $b$, and $c$) with a generic three-body coupling which we use to implement parametric couplings. For `GC' amplification between modes $a$ and $b$, we drive mode $c$, with balanced pump strengths, at both the sum (G) and difference (C) of the modes' frequencies. } 
	\label{fig:fig_1}
\end{figure*}

At present, there are two families of parametric amplifiers: those based on discrete microwave resonances, and those based on nonlinear transmission lines, so-called Traveling Wave Parametric Amplifiers (TWPAs).  TWPAs have superior instantaneous bandwidth and saturation power, but their typical reported noise performance is a factor of a few higher than resonant mode-based JPAs, and, although they are directional amplifiers, reflected pump and signal tones still often require operation with external circulators~\cite{Macklin2015, White2015, Zorin2016}.  On the other hand, resonator-based JPAs (the focus of this work) contain typically far fewer Josephson junctions, are far easier to fabricate and operate very near the quantum limit.  Their limitations are a fixed gain-bandwidth product, low input saturation powers, and lack  of directional amplification (that is, they amplify in reflection and so must be operated with external circulators)~\cite{Lehnert2007, Bergeal2010a,Hatridge2011}.  These issues stem from multiple causes. Recent work has shown that higher-order terms in the amplifier Hamiltonian can limit the saturation power of the device~\cite{Gang2017, Frattini2017, Frattini2018, Sivak2019}, while the lack of directionality and fixed gain-bandwidth product are inherent to the choice of parametric coupling used in the amplifier. However, there are numerous theoretical~\cite{Metelmann2015, Metelmann2014, Ranzani2015a} and experimental works~\cite{Sliwa2015a, Lecocq2016, Peterson2017, Fleury2014, Abdo2019} showing that applying multiple, simultaneous couplings in an amplifier containing two or more modes/ports can result in devices which potentially circumvent some or all of these limitations.

In this work, we begin by developing a kinetic-inductance shunted Josephson Ring Modulator (JRM) mixing element which is free of 4th-order nonlinearities for a certain bias flux, and embed it in a Josephson Parametric Converter (JPC)~\cite{Bergeal2010a, Bergeal2010}.   The JPC consists of three microwave modes, each linked to a single microwave port.  The JRM provides a general third order coupling between the devices three modes.  Phase preserving gain~(G) using a pair of modes is typically achieved by driving one mode off resonance at the sum of the other two modes' frequencies.  Gainless photon conversion~(C) between two modes, from which the JPC derives its' name, is achieved by driving instead at the difference frequency.

Next, we follow the scheme suggested by~\cite{Metelmann2015}, in which both G and C drives are applied simultaneously to a single pair of modes with matched coupling strengths.  The combination of frequency conjugating~(G) and non-conjugating~(C) processes results in phase-sensitive, transmission-only amplification, shown schematically in Fig.~\ref{fig:fig_1}. The device has unity reflection gain and operates bi-directionally, and so still requires external circulators, but has a number of desirable features.  First, it has a fixed, large bandwidth ($\sim$60 $\%$ of the modes' linear bandwidths). Second, rather than having divergent gain at fixed pump power, its gain is linearly dependent on pump power, with potential advantages in gain stability and device saturation behavior.  

Although this scheme has been proposed previously, efforts to implement it have been hampered by the presence of higher-order Hamiltonian terms in the JRM. Notably, unshunted~\cite{Bergeal2010} and Josephson-junction shunted JRMs~\cite{Schackert2013, Gang2017}, have 4th-order self- and/or cross-Kerr terms at all bias fluxes. These terms cause the amplifier's mode frequencies to shift with applied pump power, making biasing the GC mode of operation, which requires far more intense pumps than G or C alone, extremely difficult. 
By operating our kinetic-inductance shunted JRM at a bias point which nearly nulls all Kerr terms in the device simultaneously, we have demonstrated 21~dB of phase-sensitive gain (15~dB phase-preserving gain), with a greatly enhanced bandwidth. To demonstrate the viability of this mode of operation in practical quantum information experiments, we used the GC amplifier to measure a transmon qubit.  By strongly measuring the qubit, we were able to perform high fidelity qubit readout.  We also performed deliberately weak measurement experiments to determined the device's quantum efficiency.

\begin{figure*}[]
	\includegraphics[scale = 1]{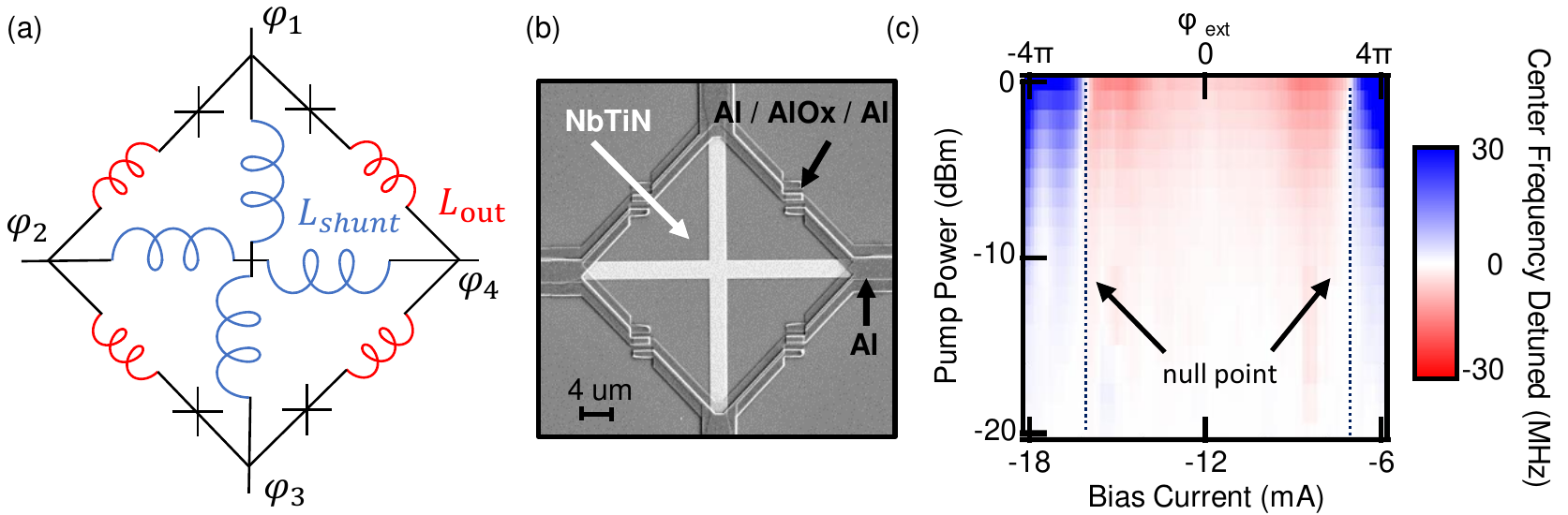}
\caption{\textbf{(a) Schematic circuit diagram of inductively shunted JRM} The circuit model for the JRM contains four Josephson junctions on each outer arm. We label the phase arocss the junctions by $\varphi_1$ to $\varphi_4$. We introduce shunted linear inductors (blue) and the stray inductors (red) to model the inductance given by the shunted wires inside the JRM and the inductance given by the wires of the outer arms. \textbf{(b) SEM image of a NbTiN shunted JRM. } Our Josephson Ring Modular features shunts made from thin ($2~\mu m$ wide by $10~nm$ thick) strips of NbTiN, which have a kinetic inductance about 5 times larger than their physical size. 
\textbf{(c) Measurement of self Kerr vs. applied flux.}  For each flux point, we represent the shift away from the small-signal resonant frequency, with positive/zero/negative Kerr with a blue/white/red color, respectively. The two cancellation points (where the plot remains white for all pump powers) are indicated with vertical dashed black lines.}
\label{fig:fig_2}
\end{figure*}

\section {Theory of the GC amplifier and the nulling of the 4th order terms}

The block diagram describing our scheme for qubit measurement and signal amplification is shown in Fig.~\ref{fig:fig_1}(a). The qubit resides inside the measurement cavity, which has two ports, a strong port and a weak port. The weak port (on the left) is used for driving the cavity and qubit while the strong port (on the right) is used for measuring the qubit. The measurement signal is routed via two circulators, which ensure that the reverse amplified quantum noise and reflected signals from the amplifier do not travel backwards to the qubit. The JPC's output is further amplified by a cryogenic HEMT amplifier before being recorded at room temperature.  The scattering matrix of the GC amplifier is shown in Fig.~\ref{fig:fig_1}(b). Signals entering a port are reflected with unity gain, and transmitted with phase-sensitive gain, whose amplified quadrature is determined by the phases of the applied pumps.

The effective circuit diagram of our inductively shunted JRM is depicted in Fig.~\ref{fig:fig_2}(a). The key ingredients of the model are (1) the Josephson junctions, (2) the shunt inductors labeled $L_{\rm shunt}$, and (3) the stray inductance on the outer arms of the JRM, labeled labeled $L_{\rm{out}}$. In this setup, the outer ring of the JRM, which contains the Josephson junctions, is the source of the non-linear couplings between the modes of the JPC. The shunt inductors are used to control the degeneracy of the ground state of the JPC at finite magnetic flux bias.  They lift the energy of states in which current flows through the shunt inductors, thus preventing the device from switching to these undesired configurations. The stray inductors are an inherent property of the aluminum traces which we use to make the outer ring of our JRM. 

The circuit is described by the Hamiltonian $H_{\rm JRM}=H_{\rm shunt}+H_{\rm out}$, composed of the shunt Hamiltonian 
\begin{align}
    H_{\rm shunt}=\sum_{i=1}^{4} \frac{\phi_0^2}{2 L_{\rm shunt}} 
    \left( \varphi_{i}-\varphi_E \right) ^2,
\end{align}
and the outer ring Hamiltonian 
\begin{align}
    H_{\rm out}=\sum_{i=1}^{4} H_{\rm seg} \left(\varphi_{i+1}-\varphi_i-\phi_{\rm ext}/4\right).
\end{align}
Here, $\varphi_i$ is the superconducting phase at the $i$-th vertex of the JRM (see Fig.~\ref{fig:fig_2}a) and we use the notation $\varphi_5=\varphi_1$~\footnote{We have assumed that the phase on the middle node is zero, this assumption is founded in the eigenmode analysis presented in Appendix~\ref{eigenmodes}.}, $\varphi_E$ is the phase at the middle point which is constrained by the outer node phase by $\varphi_E = \sum_{j=1}^{4} \varphi_j / 4$. The phase gain due to the externally applied flux bias in each of the four outer arms of the JRM is $\phi_{\rm ext}/4 = (2e/\hbar) \, \Phi_{\textrm{ext}}/4$. The energy of an outer ring segment phase biased to $\varphi$ is
\begin{align}
H_{\rm seg}(\varphi)=\min_\chi \frac{\phi_0^2}{2 L_{\rm out}} (\varphi-\chi)^{2}-E_J \cos(\chi),
\end{align}
and the minimization with respect to $\chi$ results in a transcendental equation that ensures that the current in the outer inductor is identical to the current in the Josephson junction. 

Next, we analyze the non-linear couplings between the JPC modes. Here, we focus on the case in which the JPC has a non-degenerate ground state centered on $\varphi_1=\varphi_2=\dots=0$. In this case, the phases at the JRM vertices can be expressed in terms of the canonical variables $\varphi_a$, $\varphi_b$, and $\varphi_c$ that correspond to the $a$, $b$, and $c$ eigenmodes of the JPC:
\begin{align}
&\left\{\varphi_1,\varphi_2,\varphi_3,\varphi_4\right\} \\
&=
\left\{\varphi_a + \lambda_1 \varphi_c, \varphi_b-\lambda_2 \varphi_c, -\varphi_a + \lambda_1 \varphi_c,-\varphi_b-\lambda_2 \varphi_c \right\} \nonumber.
\end{align}
The coefficients $\lambda_1=C_3/\sqrt{C_1^2 + C_3^2}$ and $\lambda_2=C_1/\sqrt{C_1^2 + C_3^2}$ come from the JPC eigemode analysis and their values depend on the capacitors added to the JRM to make the JPC, see Appendix~\ref{eigenmodes}. We obtain the non-linear couplings between the normal modes of the JPC by taking derivatives of $H_{\rm JRM}$ with respect to the canonical variables. For the self- and cross-Kerr terms we find:
\begin{align}
    K_{aa}&=\frac{1}{4!} \frac{\partial^4 H_{\rm JRM}}{\partial \varphi_a^4}=\frac{1}{6} H_{\rm seg}^{(4)}(\varphi_{\rm ext}/4)=K_{bb}=K_{cc},\\
    K_{ab}&=\frac{1}{4} \frac{\partial^4 H_{\rm JRM}}{\partial \varphi_a^2\partial \varphi_b^2}=H_{\rm seg}^{(4)}(\varphi_{\rm ext}/4)=K_{ac}=K_{bc}.
\end{align}
Observe that all of the Kerr terms are proportional to the fourth derivative of $H_{\rm seg}(\varphi)$ with respect to $\varphi$ evaluated at $\varphi=\varphi_{\rm ext}/4$.  The cosine Hamiltonian of the junction means that the Kerr naturally passes through zero, if the desired flux configuration holds to the required flux.  More, all six Kerr terms vanish identically at a single null point. In the absence of stray inductance (i.e. $L_{\rm out} \rightarrow 0$) the null  occurs at $\phi_{\rm ext}=2\pi$; the null point persists and shifts towards $\phi_{\rm ext}=4\pi$ as $L_{\rm out} \rightarrow 0$ increases (see Section~\ref{app:degeneracy}). 

We now introduce two dimensionless parameters that we will use to characterize our circuits. The first parameter, $\beta=L_J/L_{\rm shunt}=\phi_0^2/L_{\rm shunt} E_J$, measures the strength of the shunt inductors relative to the Josephson junction energy. The second parameter, $\alpha = L_{\rm out}/L_J=L_{\rm out} E_J/\phi_0^2$, measures the strength of the Josephson junction with respect to the stray inductance.  Preferably $\alpha \ll 1$ to ensure that the Josephson energy dominates the stray inductance. At the same time, $\beta$ should be sufficiently large to ensure that $\varphi_1=\varphi_2=\dots=0$ is the global minimum of $H_{\rm JRM}$ and hence the JPC has a single, well defined ground state at the null point. For the case $\alpha=0$, the non-degeneracy of the ground state for all values of $\varphi_{\rm ext}$ is ensured by setting $\beta > 4$. As $\alpha$ increases, so does the minimum required value of $\beta$ (we address these requirements by analyzing the degeneracy of the ground state in Appendix~\ref{app:degeneracy}).

\begin{figure*}[]
	\includegraphics[scale = 1]{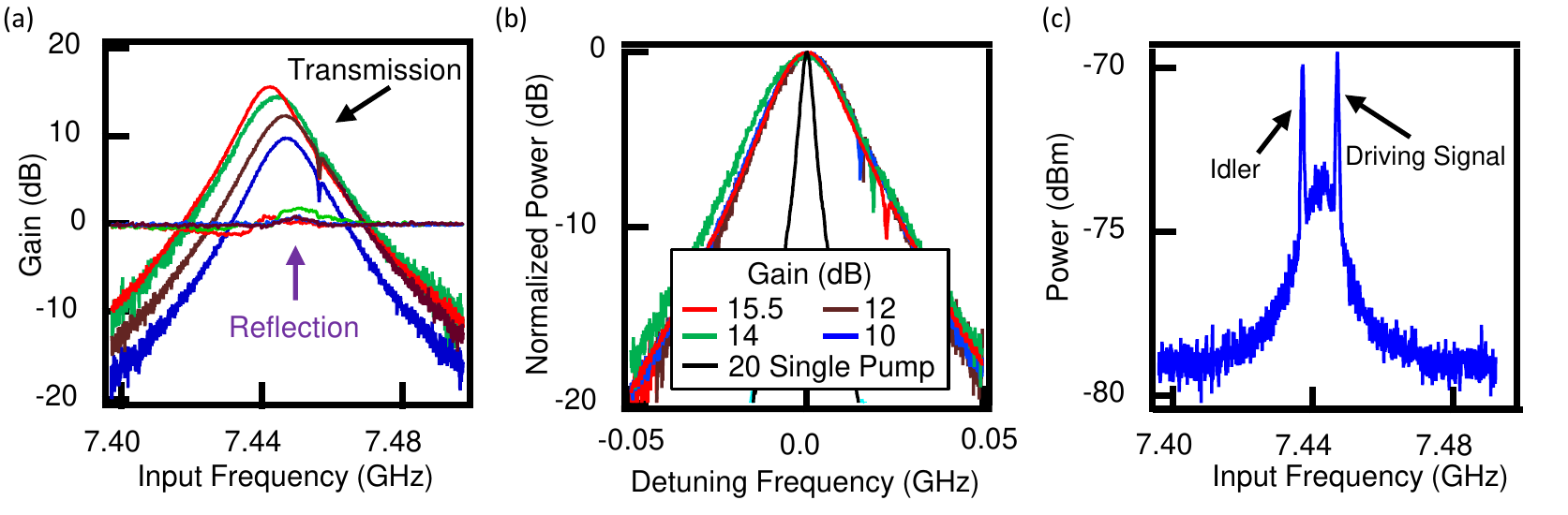}
	\caption{\textbf{ (a) Measured transmission and reflection coefficients.} For phase-preserving gains between 8 and 15 dB, we found gain in transmission, with near unity gain in reflection.   The notch at  7.45 GHz is the cavity response, which we have deliberated detuned away from for behavioral clarity.  \textbf{(b) Bandwidth comparison.} Gain curves from part $(a)$ are normalized relative to their peak gain and center frequency and are plotted together to  show the fixed bandwidth regardless of the gain. The black curve represents the gain peak and bandwidth of a 20dB, phase-preserving singly-pumped process. \textbf{(c) Phase sensitivity.} Here we show a spectrum analyzer (received power) trace when the GC amplifier is driven with a fixed tone 5~MHz detuned from its center frequency.  The presence of a symmetrically detuned idler tone below the center frequency is a clear demonstration of the phase-sensitive nature of this devices' gain. }
	\label{fig:fig_4}
\end{figure*}

A similar analysis of the JRM with Josephson-junction based shunts (as in \cite{Schackert2013, Gang2017}), shows that the cross-Kerr terms null together at a similar point to our linearly shunted JRM, but the self-Kerr null point is shifted to a larger flux.  As the behaviors of the various Kerr terms are very similar in their effect on device performance, the junction-shunted JRM ring will not realize the benefits of a linearly shunted JRM but will instead perform essentially no different from the unshunted version. We have also found, in both theory and experiment, that asymmetries in junction critical current and applied bias flux to the JRMs four loops must be minimized, as they can cause substantial difference in the flux response, and hence Kerr nulling, of the JRM.

\section{Experiment}

We form a JPC by embedding the JRM at the intersection of two, single-ended microstrip resonators, as in~\cite{Gang2017}.  The target parameters, junction $I_0=2~\mu$A $\beta >4$ and $\alpha \ll 1$, require either meandering shunt inductances~\cite{Roch2012, Flurinthesis} or kinetic-inductance shunts; in our experiments we found the latter to produce better results.  

To fabricate our JRM, we first sputter a film of 10 nm thick NbTiN onto a PMMA mask on which a cross with arm widths and lengths of  2 and 15 $\mu m$ respectively, has been patterned.  The PMMA layer is subsequently lifted off with acetone.  The resulting shunt inductors have $L_{\rm shunt} \simeq 75$ pH, about five times their geometric value. Next, standard double-angle aluminum deposition is used to create the outer ring of the JRM. As NbTiN does not form a native oxide, good contact to the upper (aluminum) portion of the circuit is achieved with only a standard, gentle Argon/Oxygen cleaning step prior to deposition of the first layer of aluminum.

In order to characterize the self- and cross-Kerr terms for a given mode/pair of modes, a pump tone is applied that is fixed 5 line widths away from the resonant frequency, while a second, weak tone is swept through the mode's resonant frequency with a Vector Network Analyzer(VNA).  The Kerr amplitude is seen as a shift in resonant frequency with increasing pump power \cite{Frattini2018}. The cancellation points at which the resonant frequency remains fixed, is shown in Fig.~\ref{fig:fig_2}(c). In subsequent qubit measurements, the frequency of the qubit cavity was tuned with an aluminum screw to align with the right most 4th-order free point of our JPC.

\begin{figure*}[]
	\includegraphics[scale = 1]{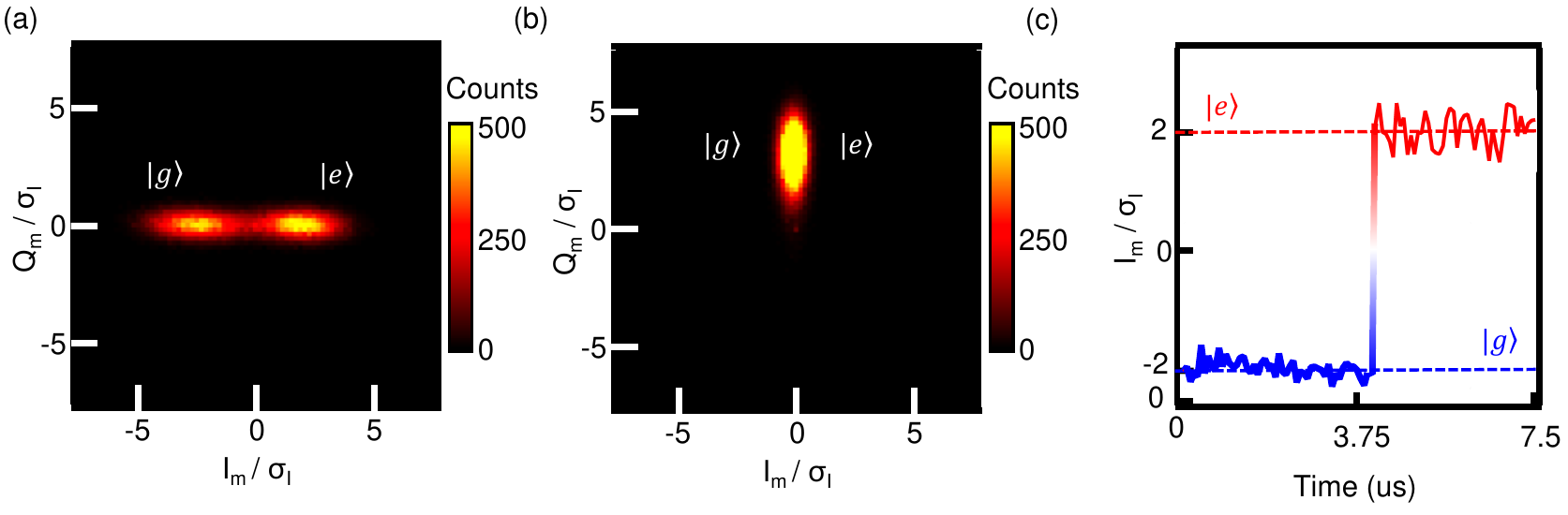}
	\caption{{\textbf{(a) Histogram of strong measurement with optimally aligned cavity drive and amplifier phase.} This histogram consists of 80,000 shots in which the qubit is prepared in the state $\ket{\Psi} = (\ket{g}+\ket{e}) / 2$ and projectively measured.  The cavity drive has been aligned to the center (phase-sensitive point) of the GC amplifier and the relative phases of the cavity drive and amplifier aligned to maximize the projectivity of the measurement. \textbf{(b) Histogram of strong measurement with orthogonally aligned cavity drive and amplifier phase.} By rotating the relative phase of cavity drive and amplifier by 90 degrees compared to part a, we show that the states of the qubit after projective measurement now completely overlap. \textbf{(c) Quantum jumps.} Here we demonstrate the ability to perform rapid QND measurements by continually monitoring the qubit and observing a well resolved quantum jump in its evolution.
	\label{fig:fig_5}
	}}\end{figure*}
	
Next, we identified the signal and idler frequencies ,which are 7.4668 GHz and 4.8715 GHz respectively, and pumped at their sum, tuning the applied pump power to achieve 20 dB gain in reflection (G). We repeated the process with the pump at the difference frequency (C), tuning the applied pump power to find a 20 dB dip in reflection. These bias powers are each very close to the critical values, and tell us the room temperature ratio of applied microwave powers required to balance the G and C processes.  We next fine tuned the applied pump frequencies until they linked to identical idler frequencies in transmission through the device.

Turning on both drives simultaneously achieves GC amplification. We control GC gain by increasing/decreasing both pumps simultaneously while maintaining the ratio established above. Measured GC amplification gain strengths between 8 and 15 dB, along with reflection performance, are shown in Fig.~\ref{fig:fig_4}(a). All transmission gain curves are measured with a VNA.  An external mixer at the difference frequency converts device outputs from the idler mode back to the input, signal mode, frequency.  The VNA is sensitive only to a single frequency, thus all gains shown are phase-preserving, and so for large gains, the phase-preserving gain peak sits $\sim$ 6~dB below the phase-sensitive gain of the device, so that the maximum phase-sensitive gain for the 15.5 dB curve (which we will use for qubit readout) is 21.5 dB.   Above 10~dB gain, the frequency of the pump tones had to be adjusted slightly to account for the imperfectly nulled higher order terms, which shifted the device's modes, and hence, the frequency of peak gain, to lower frequencies. At the same time, the reflection curves at this higher power start to show small disturbances away from unity (see Fig.~\ref{fig:fig_4}(a)). For clarity in these experiments, the amplifier is flux shifted slightly below the cavity frequency, which is evident as a  notch to the right of the gain peaks.

The other crucial feature of these gain curves is that they demonstrate the same bandwidth regardless of gain (see Fig.~\ref{fig:fig_4}(b)). We compare these bandwidths with the standard 20~dB for one single pump amplification process represented by the black curve. The GC amplifier shows 14~MHz of phase-preserving bandwidth.  In contrast, a singly-pumped gain response of 20~dB only has a bandwidth of 2.33~MHz in the same device. This bandwidth is approximately 6 times larger, in good agreement with theory.

\begin{figure*}[]
	\includegraphics[scale = 1]{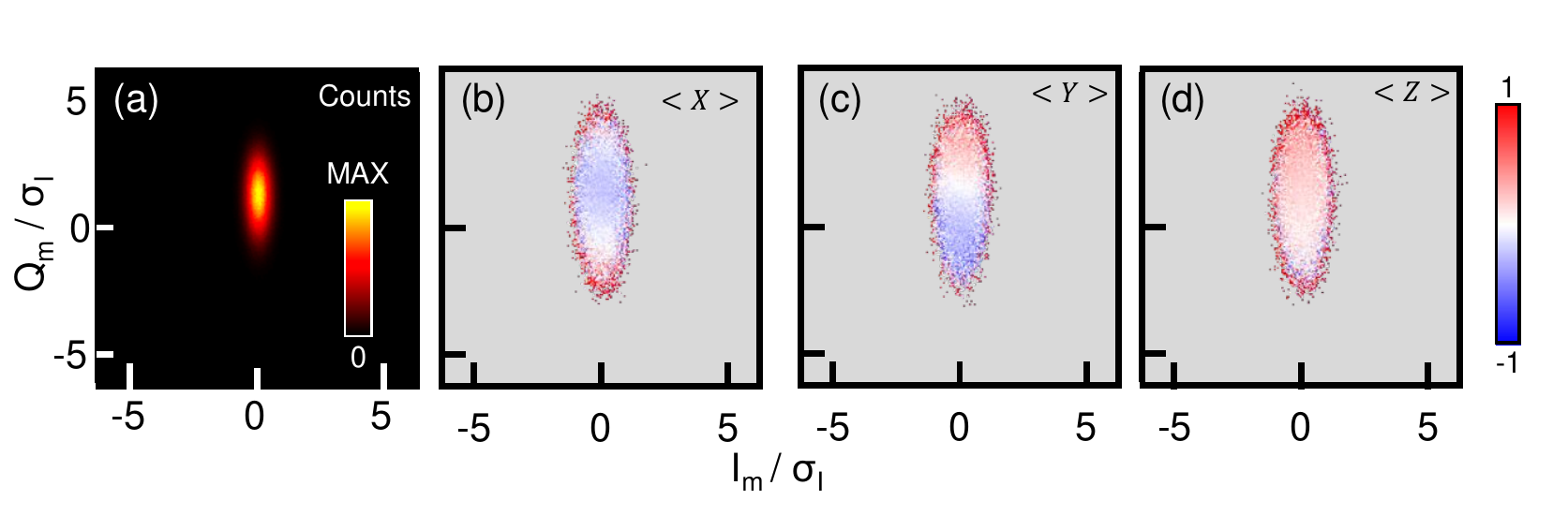}
	\caption{\textbf{(a) Histogram of weak measurement protocol with orthogonal alignment (Q-quadrature amplified) to determine quantum efficiency of the GC amplifier.} Each plot contains 80,000 measurements. \textbf{(b)-(d) Conditional expectation value of X/Y/Z after weak measurement plotted versus measurement outcome.} Sinusoidal oscillation in both X and Y are a non-classical stochastic Ramsey process. Together with the  nearly constant outcomes for Z, these plots show how an orthogonally-aligned measurement provides a `kick' around the equator of the bloch sphere.}
	\label{back_action_Q}
\end{figure*}

We also show a spectrum analyzer (received power) trace when the GC amplifier is driven with a fixed tone 5~MHz detuned from its center frequency in Fig.~\ref{fig:fig_4}(c).  The presence of a symmetrically detuned  tone below the center frequency is a clear demonstration of the phase-sensitive nature of this devices' gain.  Collectively, Fig.~\ref{fig:fig_4} demonstrate all key predicted features of GC amplification; fixed bandwidth, 0~dB gain in reflection, and phase-sensitive amplification in transmission. 

To demonstrate that our Kerr-nulled, GC pumped device is a practical, quantum-limited amplifier we also performed phase-sensitive strong/projective measurements on a superconducting transmon qubit. All further qubit measurements are performed in the configuration shown in Fig.~\ref{fig:fig_1}(a). The qubit is first prepared in the superposition state $\ket{\Psi} = (\ket{g}+\ket{e}) / \sqrt{2}$, we determined the optimal alignment  of the device's amplified quadrature by finding the largest separation between ground and excited states when projective measurement is performed, as shown in Fig.~\ref{fig:fig_5}(a).  
Rotating the relative pump phase by 180 degrees from this optimal point moves the signal to the squeezed quadrature, so that the $\ket{g}$ and $\ket{e}$ states overlap, as seen in Fig.~\ref{fig:fig_5}(b). Both histograms contain 80,000 measurements. We also measured spontaneous quantum jumps between the $\ket{g}$ and $\ket{e}$ state when the phase was rotated to the optimal alignment and the cavity was driven for 7.5~$\mu s$, as shown in Fig.~\ref{fig:fig_5}(c).  

Finally, we calculate the quantum efficiency of our amplifier via the back-action of deliberately weak measurements on the qubit's state, as in \cite{Hatridge2013, Katrinathesis}. The pulse sequence is detailed in section~\ref{PulseSequence}. For phase-sensitive amplification, only one microwave quadrature can be received at room temperature. In a qubit measurement, this confines the qubit back-action to a single plane on the Bloch sphere, determined by which quadrature is amplified.  To  calibrate our quantum efficiency ($\eta$), we  use the amplifier aligned as in Fig.~\ref{fig:fig_5}(b), resulting in a pure back-action on the qubit's phase.  We found that our quantum efficiency was $\eta=55$~\%, in good agreement with the efficiency of the device when operated as a phase-preserving amplifier at the same bias point.  We also performed weak measurements with the amplifier aligned optimally (as in Fig.~\ref{fig:fig_5}(a) (for which the back-action is only on the qubit's z-coordinate), shown in Fig.~\ref{back_action_Q}.

\section{Conclusion}

In conclusion, we have created a JPC with a linearly shunted JRM that features flux bias points where all Kerr terms can be nulled simultaneously. We used this 4th-order free JRM to realize practical, robust GC amplification, which has phase-sensitive gain in transmission, unity gain in reflection, and a fixed bandwidth that is ~6 times larger than a singly-pumped, phase-preserving 20 dB gain curve from the same device. 

We have measured a transmon qubit with this device, demonstrating high fidelity projective measurements and quantum jumps.  Finally, we used weak qubit measurements to find the quantum efficiency of this mode of amplification.  The result, $\eta=0.55$ is comparable to the efficiency of the device when operated as a singly pumped, phase-preserving amplifier.    We quote efficiency, rather than readout fidelity, as the figure of merit for the amplifier as the latter quantity involves a number of factors (such as qubit $T_1$ and insertion loss between cavity and amplifier) which are unrelated to virtues of GC amplification.  The device used in this work was also chosen to have a modest bandwidth ($\sim 20~\textrm{MHz}$), to reduce the required pump powers for GC operation.  Although the bandwidth was more than sufficient to amplify our qubit cavity, in future work we will open the bandwidth allow for multiple qubits to be read out in a single device.  

This work suggests numerous avenues for future improvements of parameteric amplifiers.  First, by extending the number of pumps to 6, while keeping the number of modes at 3, we can retain the benefits of GC pumping scheme and achieve directional amplification\cite{Metelmann2015}.  Second, we should continue to explore the wealth of potential parametric driving schemes, which may yield devices with further improved performance. Finally, by continuing to fine-tune the engineered Hamiltonian of our JRMs we can enhance the saturation power of our device\cite{Frattini2018}.
\bigbreak
\section{Acknowledgments}

C. Liu acknowledge support from a Pittsburgh Quantum Institute graduate student fellowship.  This work was supported by NSF Grant No. PIRE-1743717. Research was also sponsored by the Army Research Office and was accomplished under Grant Number W911NF-18-1-0144.  The views and conclusions contained in this document are those of the authors and should not be interpreted as representing the official policies, either expressed or implied, of the Army Research Office or the U.S. Government.  The U. S. Government is authorized to reproduce and distribute reprints for Government purposes notwithstanding any copyright notation herein.

\bibliography{Hatbibv20181113}

\appendix
\begin{widetext}
\newpage

\section{The degeneracy of the ground state of the JPC}
\label{app:degeneracy}
\begin{figure*}[b]
 	\includegraphics[width = 6.0 in]{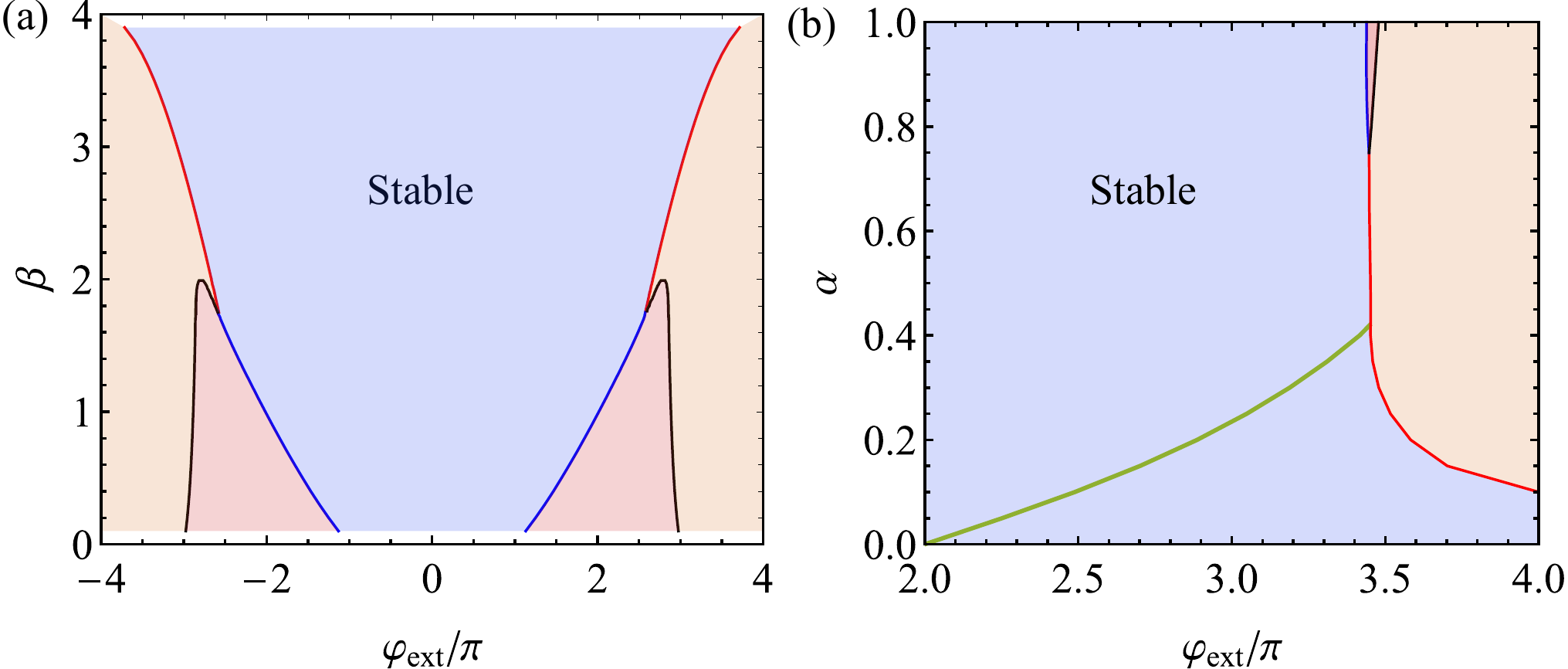}
 	\caption{ \textbf{(a) The degeneracy of the ground state without stray inductance} In this figure, We sweep the shunt parameter $\beta$ and external magnetic flux $\phi$. The blue, orange and red region show the parameter regimes that the JRM ground state is non-degenerate, two-fold degenerate and four-fold degenerate. The non-degenerate regime is considered as stable. The red, blus and black lines show the phase boundaries between stable/two-fold, stable/four-fold and two-fold/four-fold regimes. \textbf{(b) the JRM's ground state degeneracy with stray inductors.} We fix $\beta = 4.5$, and calculate the ground state degeneracy as we sweep the stray parameter $\alpha$ and external magnetic flux $\varphi_{\textrm{ext}}$. The green line in (b) shows the position of the nulling point. The nulling point shifts to higher external magnetic flux as we increase the stray parameter $\alpha$ and finally hit the unstable regime. (See discussion in the main text.)
     } 
 	\label{fig:stability}
\end{figure*}
In this section, we analyze the degeneracy of the ground states of the shunted JPC for two cases: without and with the stray inductors. Specifically, to find the degeneracy of the ground state we count the number of global minima of $H_{\rm JRM}$. 

We first consider the JPC without stray inductors, i.e. $L_{\textrm{out}} = 0$, $\alpha = 0$. We numerically find the global minima of the JRM Hamiltonian by forming a set of possible minimum points by randomly seeding a numerical minimum with initial values of $\varphi_1$, $\varphi_2$, $\varphi_3$, and $\varphi_4$. From this set, we select the minima that correspond to smallest values of $H_{\rm JRM}$, and finally we throw away repeated points. In Fig.~\ref{fig:stability}(a) we plot the degeneracy of the JRM ground state as we sweep the external magnetic flux and the shunting parameter $\beta$. The blue region in the plots corresponds to conditions under which the JRM has a single global minimum (the JPC has a non-degenerate ground state), while the orange and red regions corresponds to doubly- and quadruply-degenerate ground states.

As we decrease the inductance of the shunts, states with circulating currents move up in energy, and hence the ground state degeneracy is lifted. In other words, larger $\beta$ corresponds to less degeneracy. In the absence of stray inductors, all even order terms (beyond 2nd order) in $H_{\rm JRM}$ become zero when the external magnetic flux bias is set to $2\pi$. We find that at this flux bias the ground state is non-degenerate for $\beta > 1.0$ (minimum requirement for operation of the nulled JPC). Additionally, we find that the ground state is non-degenerate for all values of the magnetic flux bias when $\beta > 4.0$.

Next, we take stray inductance into account. The stray inductors increase the inductance of the segments of the outer ring of the JRM, effectively decreasing $\beta$ (as compared to the case $\alpha = 0$). In Fig.~\ref{fig:stability}(b) we fix $\beta = 4.5$ (close to our experimental parameters) and sweep the parameter $\alpha$ and external magnetic flux $\varphi_{\textrm{ext}}$. We notice that as we increase $\alpha$, the region of degenerate ground states reappears (near maximal magnetic flux bias). Further, we observe that as we increase $\alpha$ the Kerr nulling point moves from $\varphi_{\textrm{ext}}=2\pi$ towards $\varphi_{\textrm{ext}}=3.5\pi$, at which point it hits the degenerate ground state region (corresponding to $\alpha\approx 0.4$).

\section{The eigenmodes of the JPC} 
\label{eigenmodes}
In this subsection, we discuss the JPC circuit and identify its eigenmodes. The effective circuit diagram without input-output connections is shown in Fig.~\ref{fig:modes}(a). In order to simplify our analysis, we begin by making the assumption that $L_{\rm {shunt}} \rightarrow 0$. We will lift this assumption at the end.

\begin{figure*}[b]
    \centering
    \includegraphics[width = 6.0 in]{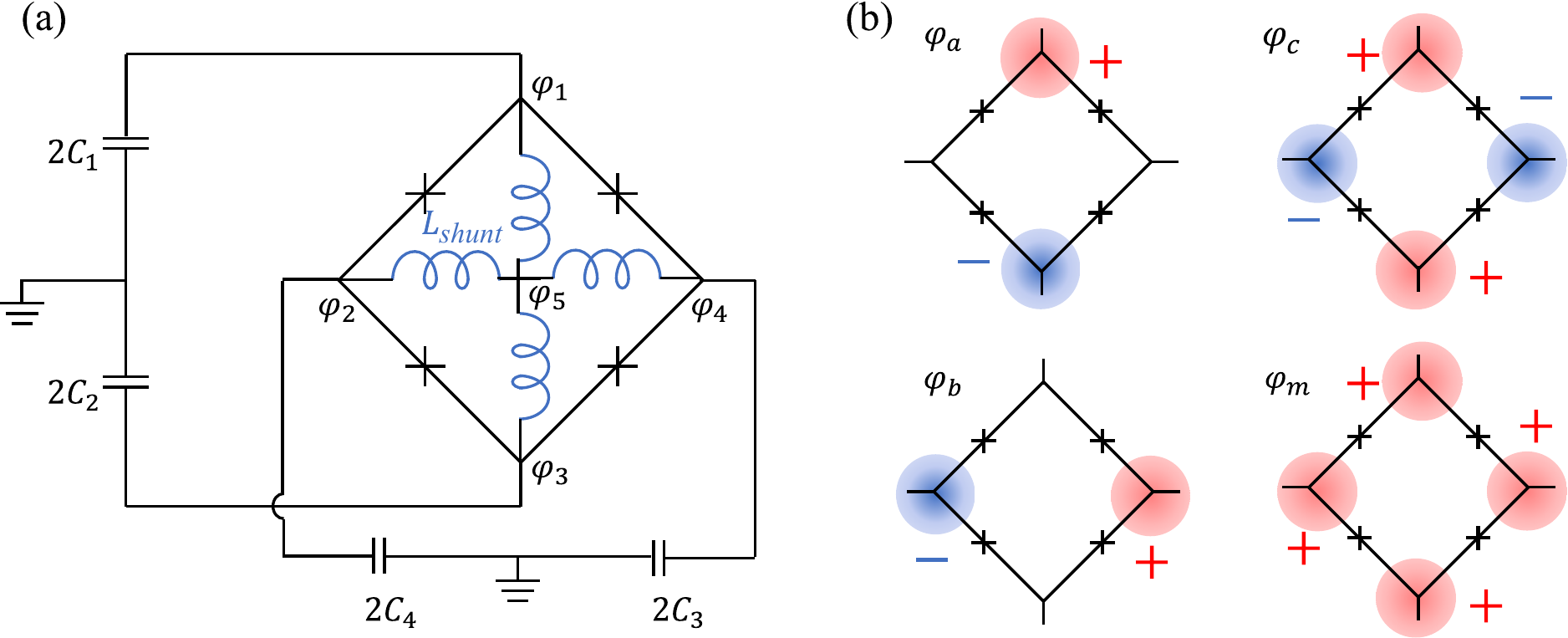}
    \caption{ \textbf{(a) The circuit diagram for the shunted JRM.} There are four normal modes of the circuit, labeled as $\varphi_a$, $\varphi_b$, $\varphi_c$ and $\varphi_m$. \textbf{(b) Engenmodes} The modes $\varphi_{a,b,c}$ are non-trivial. When the capacitance $C_1$ to $C_4$ are identical, the mode profiles are sketched.
    }
    \label{fig:modes}
\end{figure*}

In order to find the eigenmodes of the JPC, we construct the equations of motion (EOM) by applying Kirchhoff law to the circuit. We follow the standard prescription for superconducting circuits: relating voltage to the normalized flux (i.e. the superconducting phase) on each of the nodes via
 \begin{align}
     \dot{\varphi}_j = \frac{1}{\phi_0} V_j (t) \; \Leftrightarrow \; \varphi_j (t) = \frac{1}{\phi_0} \int_{-\infty}^{t}\left[ V_j(t')\right] dt' 
 \end{align}
and the supercurrent to the normalized flux via
\begin{align}
    J_j = \frac{1}{\phi_0}\frac{\partial H_{\rm JRM}}{\partial \varphi_j},
\end{align}
where $\phi_0 = \hbar / (2e)$ is the reduced magnetic flux quantum. The resulting equations of motion are 
\begin{equation}
\begin{aligned}
& 2 C_1 \phi_0^2 \ddot{\varphi}_1 + E_J \left[ \sin\left(\varphi_1-\varphi_2+\frac{\varphi_{\textrm{ext}}}{4}\right) - \sin \left( \varphi_4 - \varphi_1 + \frac{\varphi_{\textrm{ext}}}{4}\right)\right] + \frac{\phi_0^2}{L_{\textrm{shunt}}} \left( \varphi_1 - \varphi_5 \right) = 0 \\
& 2 C_4 \phi_0^2 \ddot{\varphi}_2 + E_J \left[ \sin\left(\varphi_2-\varphi_3+\frac{\varphi_{\textrm{ext}}}{4}\right) - \sin \left( \varphi_1 - \varphi_2 + \frac{\varphi_{\textrm{ext}}}{4}\right)\right] + \frac{\phi_0^2}{L_{\textrm{shunt}}} \left( \varphi_2 - \varphi_5 \right) = 0 \\
& 2 C_2 \phi_0^2 \ddot{\varphi}_3 + E_J \left[ \sin\left(\varphi_3-\varphi_4+\frac{\varphi_{\textrm{ext}}}{4}\right) - \sin \left( \varphi_2 - \varphi_3 + \frac{\varphi_{\textrm{ext}}}{4}\right)\right] + \frac{\phi_0^2}{L_{\textrm{shunt}}} \left( \varphi_3 - \varphi_5 \right) = 0 \\
& 2 C_3 \phi_0^2 \ddot{\varphi}_4 + E_J \left[ \sin\left(\varphi_4-\varphi_1+\frac{\varphi_{\textrm{ext}}}{4}\right) - \sin \left( \varphi_3 - \varphi_4 + \frac{\varphi_{\textrm{ext}}}{4}\right)\right] + \frac{\phi_0^2}{L_{\textrm{shunt}}} \left( \varphi_4 - \varphi_5 \right) = 0 \\
& \varphi_5 = \frac{1}{4} \left( \varphi_1 + \varphi_2 + \varphi_3 + \varphi_4 \right). 
\end{aligned}
\end{equation}

We can linearize the EOMs by expanding around the minimum energy configuration of the JRM (that we found in previous section). We now focusing on the non-degenerate configurations of interest: $\varphi_1=\varphi_2=\varphi_3=\varphi_4=\varphi_5=0$. In this case $\varphi_i$'s can be regarded as small perturbations away from the JRM minimum point. The resulting linearized equations of motion are
\begin{equation}
\begin{aligned}
    & \ddot{\varphi}_1 + \frac{E_J}{2 C_1 \phi_0^2} \cos(\frac{\varphi_{\rm ext}}{4}) \left[  2 \varphi_1 -  \varphi_2 - \varphi_4\right] + \frac{1}{8 C_1 L_{\textrm{shunt}}} \left(3  \varphi_1 - \varphi_2 - \varphi_3 - \varphi_4\right) =0 \\
    & \ddot{\varphi}_2 + \frac{E_J}{2 C_4 \phi_0^2} \cos(\frac{\varphi_{\rm ext}}{4}) \left[ 2 \varphi_2 -  \varphi_3 -  \varphi_1\right] + \frac{1}{8 C_4 L_{\textrm{shunt}}} \left(3  \varphi_2 - \varphi_1 - \varphi_3 - \varphi_4\right) =0 \\
    & \ddot{\varphi}_3 + \frac{E_J}{2 C_2 \phi_0^2}  \cos(\frac{\varphi_{\rm ext}}{4}) \left[ 2 \varphi_3 -  \varphi_4  -  \varphi_2 \right] + \frac{1}{8 C_2 L_{\textrm{shunt}}} \left(3  \varphi_3 - \varphi_1 - \varphi_2 - \varphi_4\right) =0 \\
    & \ddot{\varphi}_4 + \frac{E_J}{2 C_3 \phi_0^2} \cos(\frac{\varphi_{\rm ext}}{4}) \left[  2 \varphi_4 -  \varphi_1  -  \varphi_3 \right] + \frac{1}{8 C_3 L_{\textrm{shunt}}} \left(3  \varphi_4 - \varphi_1 - \varphi_2 - \varphi_3\right) =0
\end{aligned}
\label{eq:EOM_linear}
\end{equation}

Making the additional assumption that $C_1=C_2$ and $C_3=C_4$, we find that the resulting eigenvalue/eigenstate pairs are
\begin{align}
\omega_0^2&=0 & &\{1,1,1,1\} \\
\omega_a^2&=\frac{E_J}{C_1 \phi_0^2} \cos(\frac{\varphi_{\rm ext}}{4})+\frac{1}{2 C_1 L_{\textrm{shunt}}} & &\{1,0,-1,0\} \\
\omega_b^2&=\frac{E_J}{C_3 \phi_0^2} \cos(\frac{\varphi_{\rm ext}}{4})+\frac{1}{2 C_3 L_{\textrm{shunt}}} & &\{0,1,0,-1\} \\
\omega_c^2&=\left(\frac{1}{C_1}+\frac{1}{C_3}\right)
\left(\frac{E_J}{\phi_0^2} \cos(\frac{\varphi_{\rm ext}}{4})+\frac{1}{4 L_{\textrm{shunt}}}\right) & &\frac{\{C3,-C1,C3,-C1\}}{\sqrt{C_1^2+C_3^2}}
\end{align}
where $\omega_0$ is a zero mode that does not couple with the rest of the modes in the Hamiltonian. The non-trivial modes $\varphi_{a}$, $\varphi_{b}$ and $\varphi_{c}$ are sketched in Fig.~\ref{fig:modes}(b).

We now put the stray inductors back into the circuit. 
In doing so, we introduce four nodes between the stray inductors and the Josephson junctions on each arm. Because there is no kinetic terms associated with these nodes (as we neglect the capacitances to these nodes), the resulting Kirchhoff equations give us four more constrains:
\begin{equation}
    \frac{E_J}{\phi_0} \sin \left( \varphi_j - \delta_j + \frac{\varphi_{\textrm{ext}}}{4}\right) = \frac{\phi_0}{L_{\textrm{out}}} (\delta_j-\varphi_{j+1}),
    \label{eq:constraintDelta}
\end{equation}
where $\delta_j$ is the phase at the node between the stray inductor and the Josephson junction on the $j$-th arm of the JRM (see Fig.~\ref{fig:fig_2}).

In order to linearize the equations of motion, we must first find the values of $\delta_j$'s at the minimum energy configuration of the JRM. These correspond to the solution of the transcendental equation 
\begin{equation}
    \frac{E_J}{\phi_0} \sin\left(-\delta^{(0)}_j+\frac{\varphi_{\textrm{ext}}}{4}\right))  = \frac{\phi_0}{L_{\textrm{out}}}\delta^{(0)}_j.
\end{equation}
We set $\Delta=-\delta^{(0)}_j+\frac{\varphi_{\textrm{ext}}}{4}$, and use Eq.~\eqref{eq:constraintDelta} to obtain an expression for the first order correction to $\delta_j$ in terms of $\varphi_j$ and $\varphi_{j+1}$
\begin{align}
\delta^{(1)}_j=
\frac{\varphi_j \alpha  \cos(\Delta) + \varphi_{j+1}}
{1+\alpha \cos(\Delta)}.
\end{align}
We now eliminate $\delta^{(1)}_j$ from the linearized equations of motion to obtain 
\begin{equation}
    \begin{aligned}
        & \ddot{\delta \varphi}_1 
        +\frac{E_J}{2 C_1 \phi_0^2} \frac{\cos \Delta}{1+\alpha \cos \Delta} 
        \left( 2 \delta \varphi_{1} - \delta \varphi_{2} - \delta \varphi_{4} \right)
        +\frac{1}{8 C_1 L_{\textrm{shunt}}} 
        \left( 3 \delta \varphi_1 -\delta \varphi_2 -\delta \varphi_3 - \delta\varphi_4 \right) =0 \\
        & \ddot{\delta \varphi}_2 
        +\frac{E_J}{2 C_4 \phi_0^2} \frac{\cos \Delta}{1 + \alpha \cos \Delta}
        \left( 2\delta \varphi_{2} - \delta \varphi_{1} - \delta \varphi_{3} \right) 
        + \frac{1}{8 C_4 L_{\textrm{shunt}}} 
        \left(3 \delta \varphi_2 -\delta \varphi_1 -\delta \varphi_3 - \delta\varphi_4 \right) =0 \\
        & \ddot{\delta \varphi}_3 
        +\frac{E_J}{2 C_2 \phi_0^2} \frac{\cos \Delta}{1 + \alpha \cos \Delta}
        \left( 2\delta \varphi_{3} - \delta \varphi_{2} - \delta \varphi_{4} \right) 
        + \frac{1}{8 C_2 L_{\textrm{shunt}}} 
        \left(3 \delta \varphi_3 -\delta \varphi_1 -\delta \varphi_2 - \delta\varphi_4 \right) =0 \\
        & \ddot{\delta \varphi}_D 
        +\frac{E_J}{2 C_3 \phi_0^2} \frac{\cos \Delta}{1 + \alpha \cos \Delta}
        \left( 2\delta \varphi_{4} - \delta \varphi_{1} - \delta \varphi_{3} \right) 
        + \frac{1}{8 C_3 L_{\textrm{shunt}}} 
        \left(3 \delta \varphi_4 -\delta \varphi_1 -\delta \varphi_2 - \delta\varphi_3 \right) =0
    \end{aligned}
    \label{eq:stray_JRM_EOM_linear}
\end{equation}
Comparing equations Eq.~\eqref{eq:EOM_linear} and Eq.~\eqref{eq:stray_JRM_EOM_linear} we observe that the effect of stray inductors is a renormalization of the Josephson energy $E_J \rightarrow E_J \frac{\cos \Delta}{1 + \alpha \cos \Delta}$. This renormalization results in a shift of the eigenfrequencies but the eigenmodes remain identical.

\section{Measurement of Kerr terms}

\begin{figure*}[]
	\includegraphics[scale = 0.85]{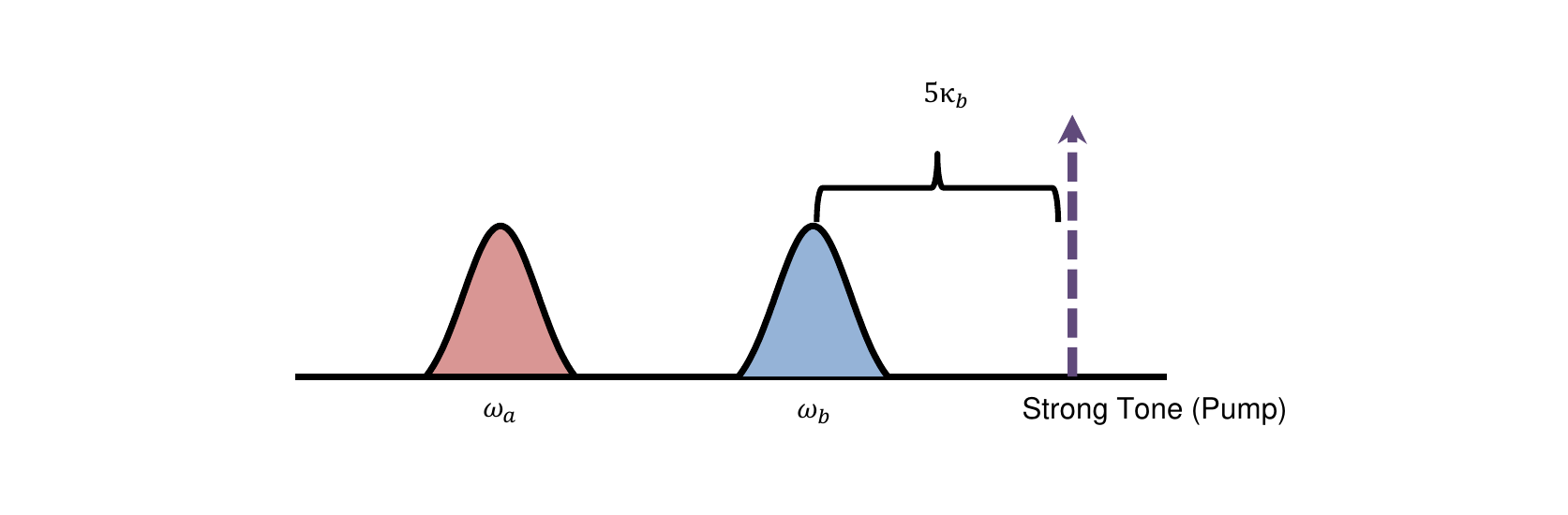}
	\caption{\textbf{Visualization of how to measure cross Kerr.} We prepare a pump tone that is 5 times linewidth away from the mode frequency. For instance, in the figure I prepare a pump tone that is 5 linewidth away from mode $b$ and observe frequency shift on mode $a$. The color plot is the result of self-kerr of mode a. Color represents how much frequency does mode $a$ shift away compared to that of low power. As the result, the white regime indicates the mode frequency remain the same while increasing of pump power. The fourth order free point will be the place where white strips appear.}
	\label{Kerr_measure}
\end{figure*}

We need to quantify the strength of the Kerr terms inherent to our amplifier in order to find when they go to zero. To do this, we perform Self-Kerr and Cross-Kerr duffing sweeps across a wide range of flux. The self Kerr terms, $a^\dagger a a^\dagger a$ and $b^\dagger b b^\dagger b$, can be measured by applying a strong pump tone 5 line widths (here 100 MHz from the signal and 125 MHz  from the idler) away from the respective resonant frequencies. We then sweep the flux and the pump power and measure the detuning from the resonant frequencies. Plotting this detuning visually in red, for negative detuning, and blue, for positive detuning, allows us to quickly identify where the these terms go to zero, since zero detuning, representing no 4th-order effects, is plotted in white. The cross Kerr, $a^\dagger a b^\dagger b$  is measured similarly, but this time we apply a tone 5 line widths away from the signal mode and measure the effect of the idler, as seen in Fig. \ref{Kerr_measure}. With a high enough pump power, it becomes easy to distinguish the red/blue regions, and thus, the crossing zone in the middle where the specific 4th-order term goes to zero. This flux bias point as shown Fig. \ref{fig:fig_2}~(c), is represented by a dotted black line. This JRM is designed such that all 4th order terms should be identically nulled at the same flux bias.

\section{Back action on the qubit for weak amplification}
\label{PulseSequence}

In addition to being an essential component to the measurement chain used to readout superconducting qubits, parametric amplification can be used to manipulate them as well. For instance, the back-action associated with parametric amplification can perform the essential function of remote entangling distant qubits. It can also  be used as an accurate, self-calibrating way to determine measurement efficiency. These types of measurements will be vital in identifying and eliminating the effects which limit our ability to manipulate quantum systems.

\begin{figure*}[h!]
	\includegraphics[scale = 1]{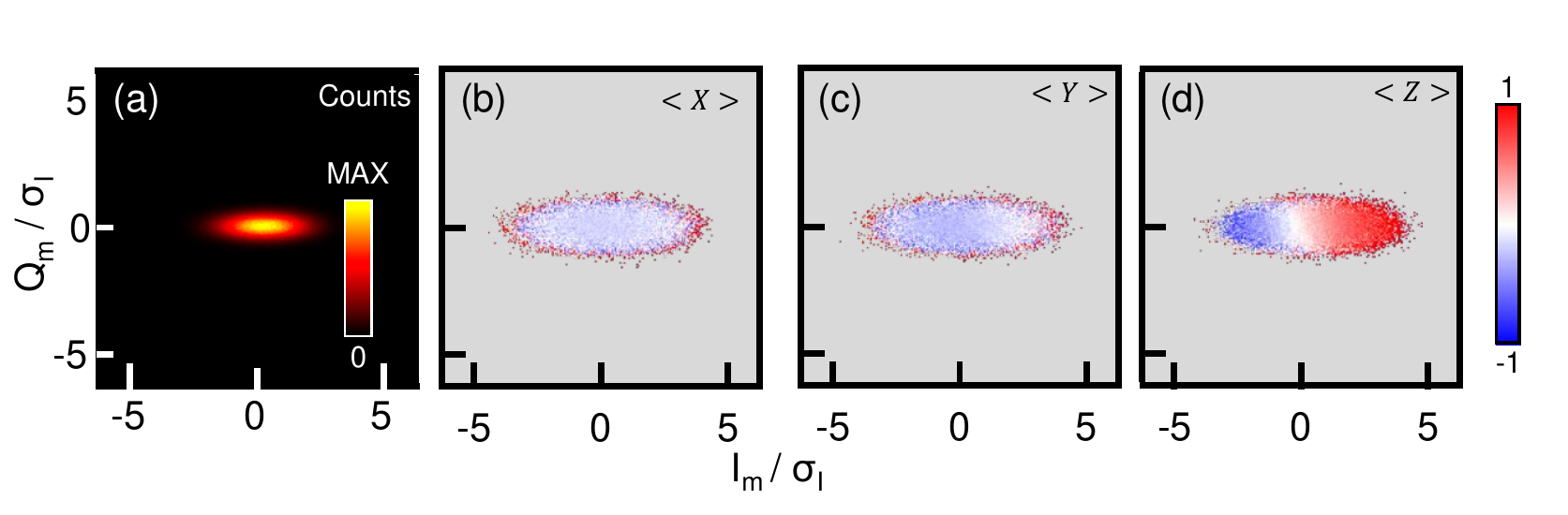}
	\caption{\textbf{(a) Histogram of weak measurement protocol with optimal alignment (I-quadrature amplified) to determine quantum efficiency of the GC amplifier.} Each plot contains 80,000 measurements. \textbf{(b)-(d) Conditional expectation value of X/Y/Z after weak measurement plotted versus measurement outcome.} These plots show how an optimally-aligned measurement provides a `kick' around the Y-Z plane of the bloch sphere.}
	\label{fig:fig_6}
\end{figure*}

In Fig.~\ref{fig:fig_6}, we apply second measurement strength to be 5 times (in voltage) smaller than that of strong measurement while GC is operated at orthogonal phase. The histogram shows sinusoidal oscillation in both X and Y o highly nonclassical stochastic Ramsey processes. The expectation value of Z is affected by qubit's relaxation and leads to the slightly red color instead of complete white. With the same procedure but squeezing the Q quadrature instead, we change's qubit's motion to be confined to the  Y-Z plane, see Fig.~\ref{fig:fig_6}. 

The standard way to perform back-action measurement is shown in Fig.\ref{back_action}. We first strongly read out the qubit and record the outcome, which will be used  to prepare the qubit in the ground state with state selection. Then the qubit is rotated by the +y axis and measured with a variable measurement strength, and the outcome is recorded again. For the final tomography, phase measures the x, y, or z component of the qubit with a strong measurement pulse. 

If we take the qubit to originally be oriented along the Y-axis and the I quadrature to be perfectly squeezed,  the back-action corresponds to stochastic, trackable motion of the qubit state in the x-y plane, with the extent of the motion varying with the strength of the measurement.  For weak strength, the back-action looks like a stochastic rotation in the x-y plane with the degree of rotation encoded in Qm.

\begin{figure*}[h!]
	\includegraphics[scale = 1]{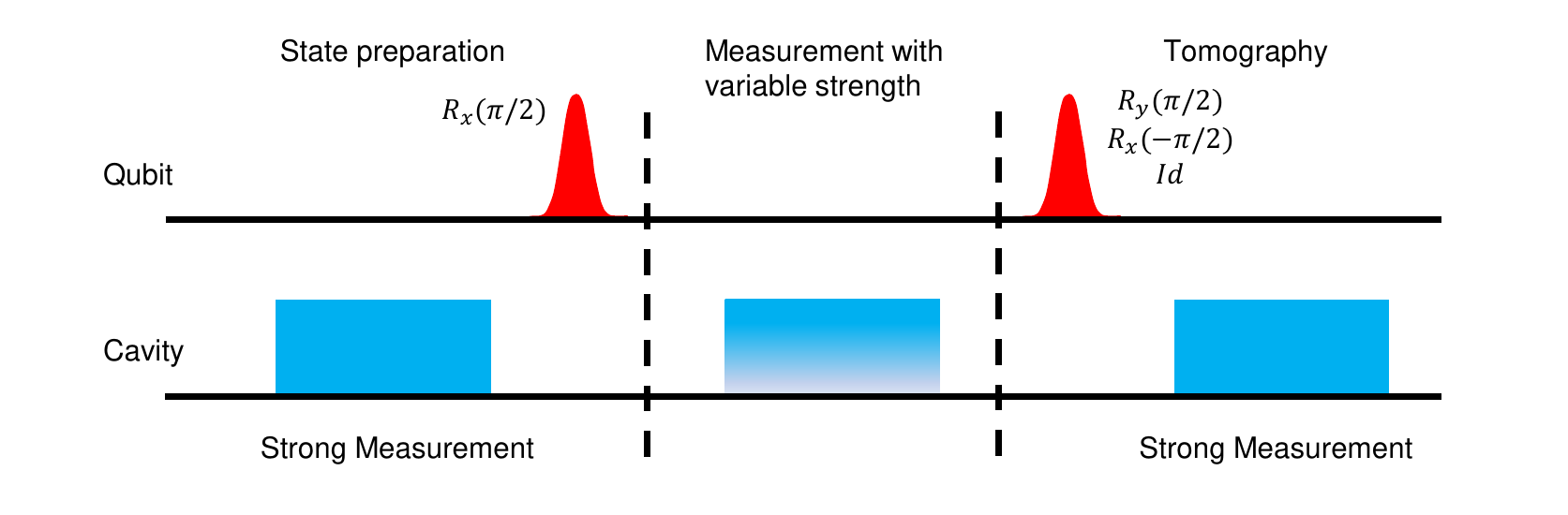}
	\caption{{\textbf{Pulse sequence for quantifying measurement back-action.} We first strongly read out the qubit and record the outcome, which will be used to prepare the qubit in the ground state by postselection. Then, the qubit is rotated by the +y axis and measured with a variable measurement strength, and the outcome is recorded. The final tomographs, phase measures  the x, y, or z component of the qubit Bloch vector with a strong measurement pulse.}}
	\label{back_action}
\end{figure*}

\begin{figure*}[]
	\includegraphics[scale = 1]{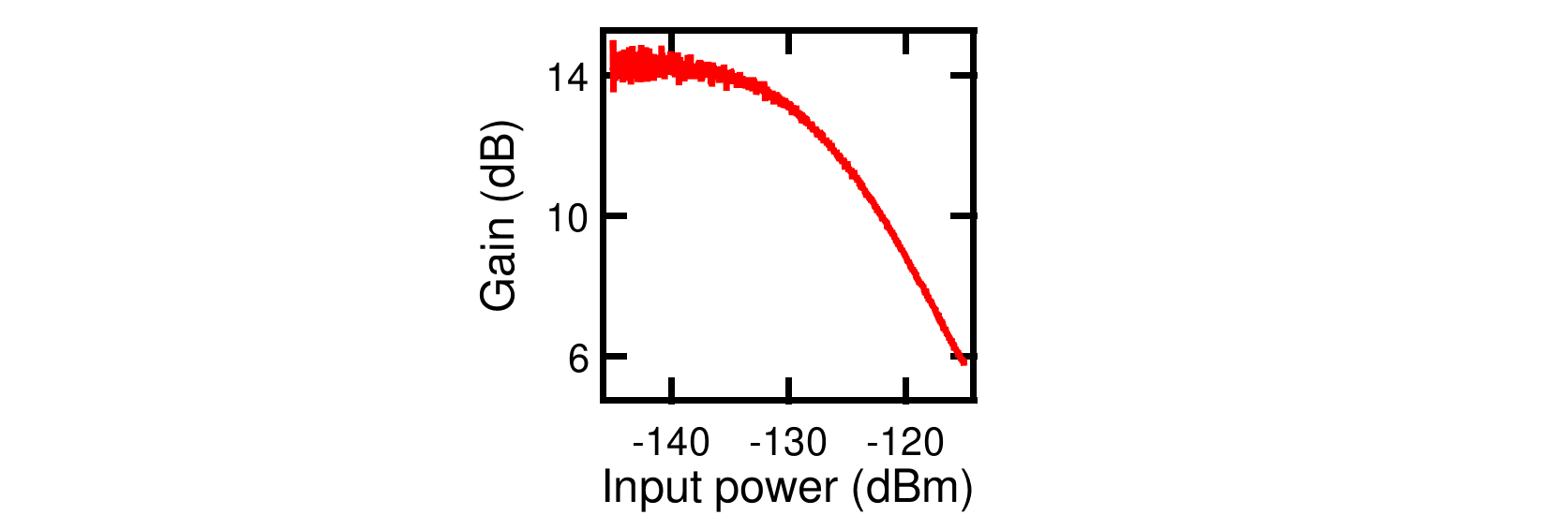}
	\caption{\textbf{Saturation power.} The saturation power of our GC device at around 14~dB is measured. $P_{\textrm{-1dB}}$  is around -130~dBm.}
	\label{saturation}
\end{figure*}

\section{Saturation power}
The amplifier's saturation behavior was measured as shown in Fig.~\ref{saturation}.  Although operated with much higher pump powers, this dynamic range is virtually identical to singly-pumped phase-preserving and phase-sensitive gain at the same Kerr nulling point.  This points to conventional pump depletion not limiting the device.  In a separate work, we have concluded that still higher-order nonlinearities can contribute strongly to gain saturation and must be controlled to realize amplifiers with superior dynamic range \cite{chenxu2019}.

\end{widetext}

\end{document}